# Canalization acoustic phonon polaritons in metal-MoO$_3$-metal sandwiched structures for nano-light guiding and manipulation

Qizhi Yan,† Runkun Chen,† Zhu Yuan, Peining Li,* and Xinliang Zhang*

We theoretically propose and study in-plane anisotropic acoustic phonon polaritons (APhPs) based on a layered structure consisting of a monolayer (or few layers) α-phase molybdenum trioxide (α-MoO$_3$) sandwiched between two metal layers. We find that the APhPs in the proposed sandwiched structures are a canalization (highly directional) electromagnetic mode propagating along with the layers and at the same time exhibit extreme electromagnetic-field confinement surpassing any other type of phonon-polariton modes. When a double layer of α-MoO$_3$ is sandwiched by two Au layers, twisting the two α-MoO$_3$ layers can adjust the interlayer polaritonic coupling and thus manipulate the in-plane propagation of the highly confined APhPs. Our results illustrate that the metal-MoO$_3$-metal sandwiched structures are a promising platform for light guiding and manipulation at ultimate scale.

## Introduction

Phonon polaritons (PhPs) in polar van der Waals (vdW) layered materials (e.g., hBN, α-MoO$_3$, and α-V$_2$O$_5$) enable large electromagnetic-field confinement[1–6], long lifetime[7–9], hyperbolic dispersion[10–14], and extremely slow group velocities[9,15–17] for realizing strong light-matter interactions at nanoscale[18–20]. These unique properties facilitate various applications, such as vibrational strong coupling[19,21,22], high-quality nanoresonators[23–25], enhanced mid-infrared light detection[12,14,26], among others[27,28]. Importantly, the layer thickness of vdW materials can be precisely controlled with a one-atomic-layer resolution by simple mechanical exfoliation, yielding the fine-tuning of the dispersion (and thus the confinement) of PhPs in thin layers of polar vdW materials[29–31]. Particularly, when reducing the thickness to an one-atomic layer, a large confinement factor of $k_p/k_0 = 60$ was experimentally demonstrated for PhPs in a monolayer hBN ($k_p$ and $k_0$ being the wavevectors of PhPs and free-space photons, respectively)[31].

We have recently proposed the concept of acoustic phonon polaritons (APhPs) to enhance further the polariton confinement, based on a monolayer hBN located over a metal layer by a separation gap[18]. Similarly to the surface plasmons in graphene/metal heterostructures[32–34], the proposed APhPs modes originate from the strong near-field coupling of PhPs in the hBN monolayer and their electromagnetic "mirror images" in the metal[35]. For that reason, APhPs are also called *image polaritons*. As the result of the near-field coupling, decreasing the gap width to a few nanometers can significantly increase near-field confinement and enhancement of APhPs in the hBN/metal heterostructure, which providing a promising platform for studying light-matter interactions. However, hBN-based APhPs modes show isotropic in-plane propagation along the layers, exhibiting the wavevectors limited by a closed circular isofrequency contour in the $k$-domain. Nonetheless, one still expects a limit for further increasing the confinement of APhPs when reducing the gap size to zero.

As a prototypical biaxial polar vdW crystal, α-MoO$_3$ supports the PhPs exhibiting in-plane hyperbolic dispersion and an open, hyperbolic isofrequency diagram in $k$-domain which hold a huge wavevector that can realize a stronger confinement than isotropic PhPs[26,36–43]. Recently, APhPs in MoO$_3$/metal structure have been studied, which hold a great enhancement in the light–matter interaction and provides a theoretical platform for the detection of a single molecule[44]. Here we propose and theoretically explore in-plane anisotropic APhPs based on metal/MoO$_3$/metal sandwiched structures which can yield in-plane hyperbolic APhPs that achieve stronger electromagnetic confinement than anisotropic APhPs in MoO$_3$/metal structure. Interestingly, the APhPs in the sandwiched structure are highly confined, canalization (highly directional, diffraction-less) waveguide modes propagating along with the layers, possessing excellent capabilities for nanoscale light guiding and manipulation. We further show that the hyperbolic APhPs in a double layer of α-MoO$_3$ between the two metal layers can be tuned by twisting the angle $φ$ between the optical axes of the two separated layers of α-MoO$_3$ and realize a topological transition from open to closed dispersion contours[45–48].

## Methods and results

To begin with, we perform numerical simulations to calculate the electric field distribution of APhPs using the software package COMSOL. We first compare two types of APhPs modes: the in-

Wuhan National Laboratory for Optoelectronics & School of Optical and Electronic Information, Huazhong University of Science and Technology, Wuhan 430074, China

E-mail: lipn@hust.edu.cn, xlzhang@mail.hust.edu.cn

†These authors contributed equally to this work.

plane isotropic APhPs (Fig. 1a to c) and the in-plane hyperbolic APhPs (Fig. 1d to f). To provide a more proper comparison, we try to use the most similar material parameters in the simulations for modeling those two different modes. For simulating the hyperbolic APhPs, we model a monolayer α-MoO$_3$ as a two-dimensional (2D) conductive sheet[13,31] with in-plane anisotropic conductivities ($\sigma_{x,MoO3} \neq \sigma_{y,MoO3}$, parameters according to Ref.49). The monolayer α-MoO$_3$ locates at the height of 0.5 nm above an Au layer (Fig. 1d). On the other hand, we virtually set a conductive sheet with isotropic conductivities ($\sigma_x = \sigma_y = \sigma_{x,MoO3}$) for calculating the isotropic APhPs, which also locates at a distance of 0.5 nm above the Au substrate (Fig. 1a). To excite the polaritons, we place a vertical electric dipole source at the height of 0.5 nm above the 2D sheets. We also consider nonlocal Au layers underneath 2D conductive sheets to account for the extremely large polariton wavevectors[18,50].

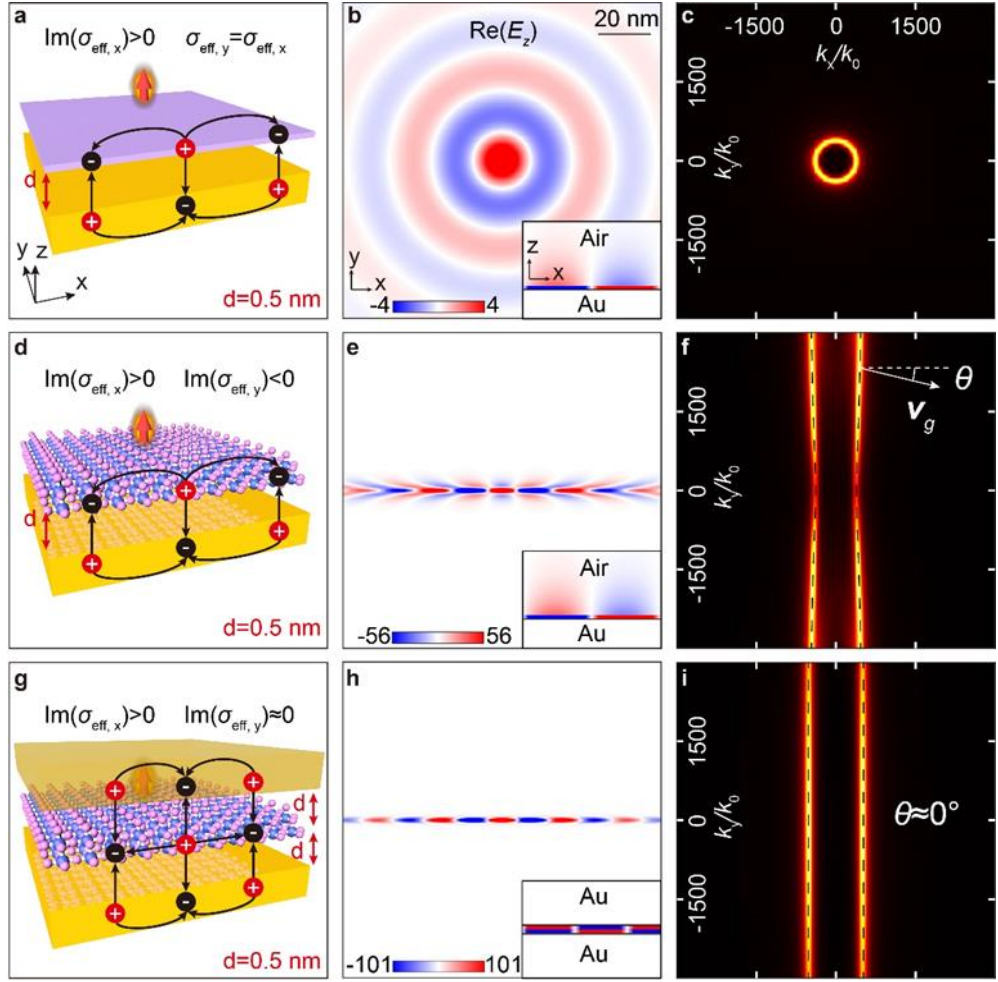

**Fig. 1**. Dipole-launching of acoustic phonon polaritons (APhPs) on three different structures. (**a**, **d** and **g**) Schematics of the dipole-launching of APhPs on three different structures: (a) the isotropic heterostructure; (d) the α-MoO$_3$/Au heterostructure; (g) the Au/α-MoO$_3$/Au sandwiched structure, respectively. In all cases, the dipole source locates at the height of 0.5 nm above the α-MoO$_3$. The width of the air gap is $d = 0.5$ nm. (**b**, **e** and **h**) Simulated near-field distribution (the component Re($E_z$)) above the Au surface for those three structures at the frequency $\omega = 830$ cm$^{-1}$, respectively. The inset in each figure shows the cross-section of the electric field Re($E_z$) from the $x$-$z$ plane. (**c**, **f** and **i**) Absolute value of the Fourier transform (FT) of the panel (b), (e) and (h), respectively. The dashed black lines are the theoretical dispersion calculated using a transfer matrix method. We define the angle $\theta$ between the group velocity $\boldsymbol{v}_g$ and the $k_x$-axis. The divergence angle $\theta$ in (i) is almost zero.

Figure 1b displays the electric-field distribution of isotropic APhPs (taken at the plane above Au surface) at the frequency of $\omega = 830$ cm$^{-1}$, which propagating with circular wavefronts and exhibiting a closed circular distribution (radius $k = 380k_0$) in $k$-space (Fig. 1c). In contrast, the hyperbolic APhPs presented in Fig. 1e show highly directional, strongly anisotropic in-plane propagation. Their in-plane wavevectors form an open wide, hyperbolic isofrequency contour in $k$-space (Fig. 1f), in excellent agreement with the theoretical hyperbolic dispersion (dashed black line) calculated using a transfer matrix method[51]. Remarkably, the wavevector components $k_y$ of hyperbolic APhPs are significantly larger than that of isotropic APhPs. It verifies that the in-pane hyperbolic APhPs can provide much stronger electromagnetic confinement in real space (particularly in the $y$-direction) than the in-plane isotropic APhPs. The hyperbolic APhPs also exhibit much stronger near-field enhancement in the air gap between α-MoO$_3$ and the Au layer (see the color bars and the inserts in Fig. 1b and e) than isotropic APhPs.

Considering that the electromagnetic-field concentration (confinement and enhancement) of APhPs arises from the coupling of the α-MoO$_3$ sheet and the Au layer, we come up with the idea of using a sandwiched structure to enhance near-field interactions. As sketched in Fig. 1g, we consider two Au layers located symmetrically above and underneath the α-MoO$_3$ sheet (with two air gaps: $d = 0.5$ nm). The dipole source locates at the height of 0.5 nm above the 2D α-MoO$_3$ sheets to excite the polaritons. Figure 1h shows the simulated electric field distribution in the sandwiched structure, taken from the plane at the surface of the lower Au layer. Compared to the heterostructure (Fig. 1e), the APhPs in the sandwiched structure are more directional, behaving as a canalization (directional, diffraction-less) waveguide mode. The isofrequency contour of the polariton wavevectors is almost flat (Fig. 1i), yielding that the divergence angle $\theta$ (between the group velocity $\mathbf{v}_g$ and the $k_x$-axis) is nearly zero and the maximum $k_y$ even reaches more than $3000k_0$ (thus extreme spatial confinement < 5 nm in the $y$-direction). At the same time, the near-field enhancement inside the air gap of the sandwiched structure (Fig. 1h, the color band, and the insert) is further increased to be almost two times compared to the heterostructure. Our results provide a simple way to realize a canalization APhPs waveguide mode than the twisted stacked α-MoO$_3$ layers structure, which hold a promising platform for ultimate-scale light guiding and manipulation.

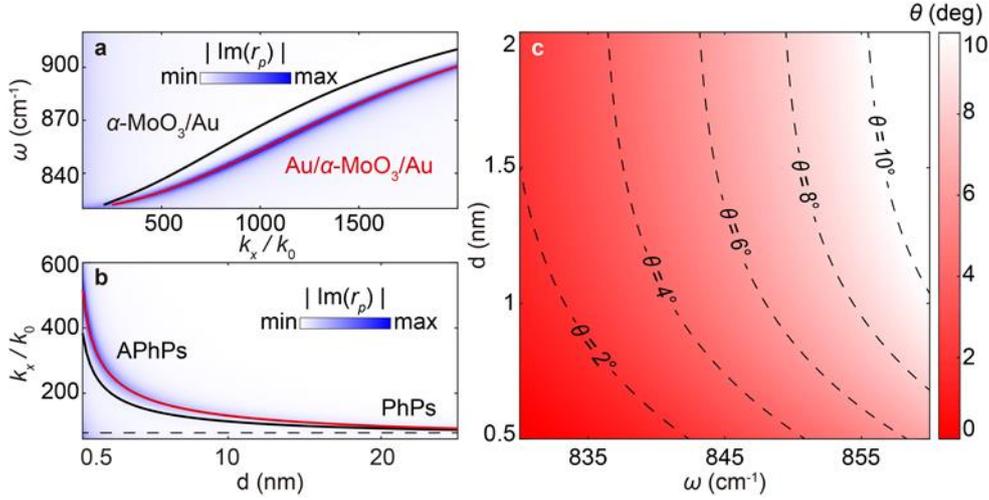

**Fig. 2.** Tuning the APhPs dispersion and the angle $\theta$ with changing the width of the air gap. (**a**) Dispersions of the APhPs modes (propagating along the $x$-axis) in the Au/α-MoO$_3$/Au structure (red line) and the α-MoO$_3$/Au structure (black line) for the width of each air gap $d = 0.5$ nm, respectively. False-color map depicts the absolute value of the imaginary part of the Fresnel reflection coefficient $|\text{Im}(r_p(\omega, k))|$ for the sandwiched structure as a function of the frequency and the wavevector ($d = 0.5$ nm). (**b**) Variations of the wavevector of the APhPs modes (propagating along the $x$-axis) with changing the width $d$ for the Au/α-MoO$_3$/Au structure (red line) and the α-MoO$_3$/Au structure (black line), respectively. False-color map shows the $|\text{Im}(r_p(d, k))|$ for the sandwiched structure as a function of the air gap and the wavevector at the fixed frequency of $\omega = 830$ cm$^{-1}$. (**c**) False-color map shows the dependence of the divergence angle $\theta(\omega, d)$ of APhPs in the sandwiched structure.

To further compare the polariton confinement, we calculate the dispersion of the APhPs mode propagating along the $x$-direction for both the Au/α-MoO$_3$/Au structure and the α-MoO$_3$/Au structure. The false-color map shown in Fig. 2a depicts the absolute value of the imaginary part of the Fresnel reflection coefficient for the sandwiched structure as a function of the frequency and the wavevector, $|\text{Im}(r_p(\omega, k))|$. The peak observed in the map (indicated by a red line) corresponds to the APhPs mode in the sandwiched structure. For comparison, the black line in Fig. 2a shows the dispersion of APhPs in the α-MoO$_3$/Au structure. For a given frequency, the wavevectors of the former case are larger than that of the latter, verifying the increased electromagnetic-field confinement provided by the sandwiched structure.

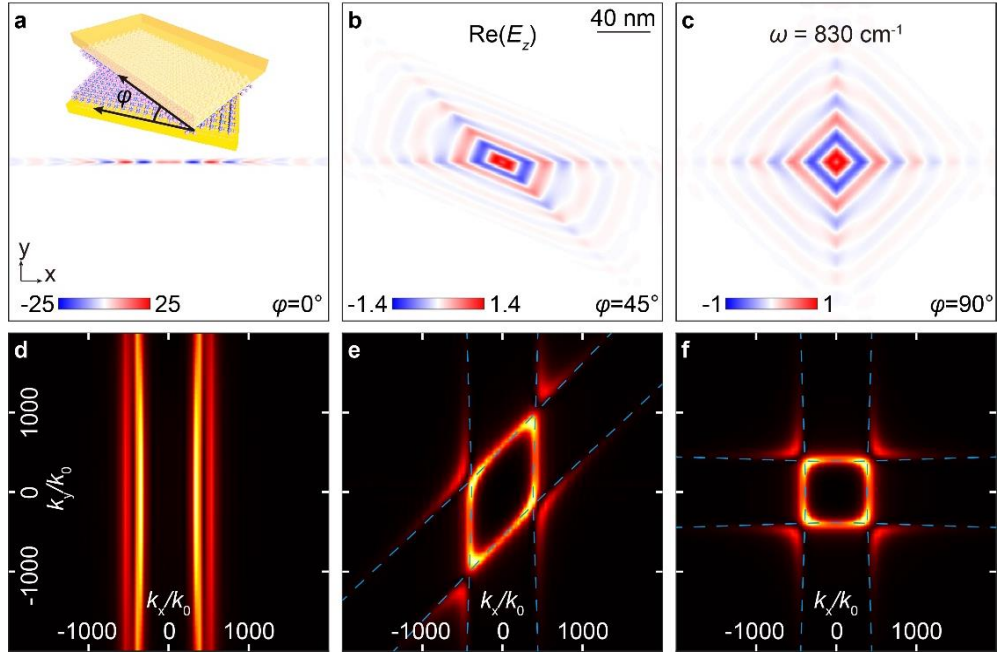

**Fig. 3.** Configurable APhPs in the twisted structure consisting of a double layer of α-MoO₃ sandwiched by two Au layers. (**a-c**) Simulated electric field distributions (the component Re($E_z$)) of the dipole- launched APhPs in the sandwiched structure (d=0.5 nm) for three different twist angles: $\varphi$ = 0° (a), 45° (b) and 90° (c). (**d-f**) Corresponding FT of the images shown in (a-c), showing the bright peaks that describe the isofrequency curves of the polaritons in the twisted structure. On the other hand, we also mark the wavevector isofrequency curves ( dashed blue lines) of the individual polariton modes in each single isolated layer (without considering the coupling), respectively. It is seen that the polaritonic coupling significantly modify the isofrequency curves of the polariton wavevector in the twisted structure.

The electromagnetic-field concentration of APhPs is also controlled by the width of the air gap. To show the influence of the gap width $d$, in Fig. 2b, we show the Fresnel reflection coefficient of the sandwiched structure as a function of $d$ and $k$ for the fixed frequency $\omega = 830$ cm$^{-1}$. We can see that the APhPs wavevector (indicated by a red line) increases with decreasing $d$. It achieves $k = 508k_0$ when $d = 0.5$ nm, about 34% larger than APhPs in the heterostructure ($k = 380k_0$). When increasing the gap width (the coupling thus weakened), the wavevectors of both APhPs modes decrease dramatically and approach that of the PhPs in the monolayer α-MoO₃. In addition to the electromagnetic-field concentration, another important quantity is the divergence angle $\theta$, from which one can evaluate the canalization propagation. Figure 2c shows a false-color map describing the divergence angle $\theta(\omega, d)$ of the APhPs in the Au/α-MoO₃/Au structure. We can see that, to obtain the canalization APhPs (here defined for $\theta < 2°$), it requires $d < 1.5$ nm and $\omega < 840$ cm$^{-1}$.

We also further explore the APhPs in a double layer of α-MoO₃ sandwiched by two Au layers at the frequency of $\omega = 830$ cm$^{-1}$, as sketched in Fig. 3a, the air gaps between α-MoO₃ with Au layer and between α-MoO₃ with α-MoO₃ are 0.5 nm and 1 nm respectively. This scheme has an obvious advantage that the polariton propagation can be adjusted by controlling the twist angle $\varphi$ between the optical axes of the two separated layers of α-MoO₃. For $\varphi = 0°$, the APhPs exhibit anisotropic canalization propagation (Fig. 3a), similar to Fig. 1h. However, the Fourier transform (FT) of Fig. 3a reveals two flat isofrequency contours in $k$-space (Fig. 3d), indicating the existence of two polariton modes: symmetric (with larger $k$) and antisymmetric (with lower $k$) modes. They arise from the hybridization of the polariton mode (dashed blue line) in each layer of α-MoO₃. As shown in Fig. 3a to c, when increasing $\varphi$ from 0° to 90° (thus changing the polaritonic coupling between the two layers), the polaritons are tuned to exhibit more isotropic propagation, showing such as the parallelogram wavefronts for $\varphi=45°$ (Fig. 3b and e) and the square wavefronts for $\varphi=90°$ (Fig. 3c and f), realizing a topological transition from open to closed dispersion contours. These results present the polaritonic coupling at a deeply subwavelength scale (the wavevector $k > 500k_0$), holding a great promise for realizing tunable nanoscale light-matter interactions.

## Scheme for experimental demonstration

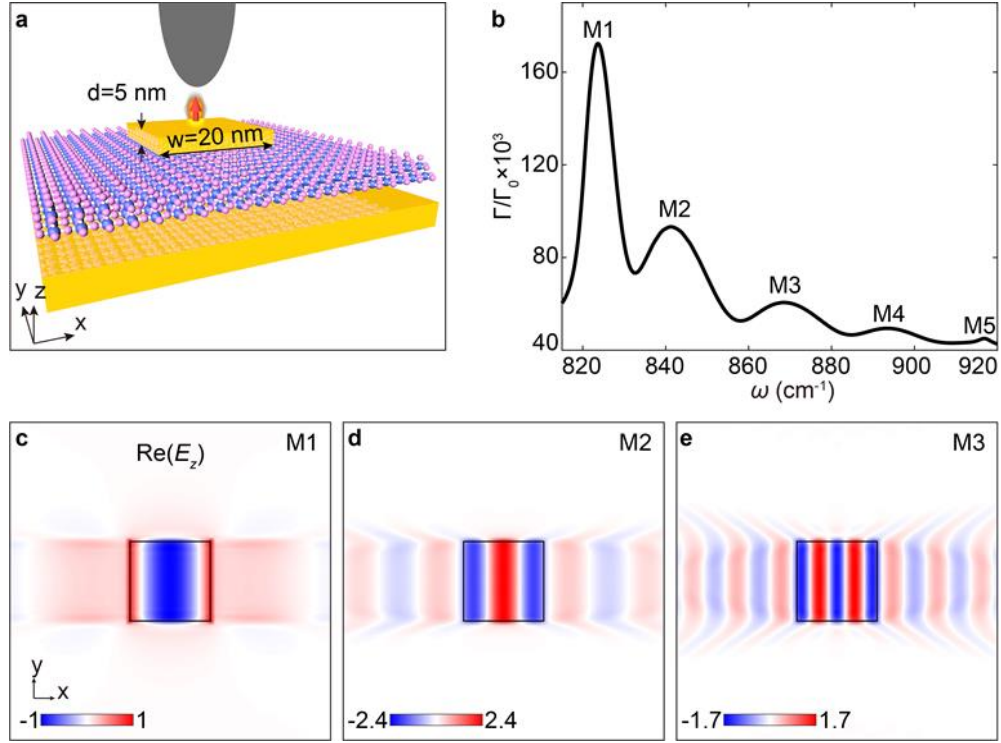

**Fig. 4.** Theoretical proposals for the experimental verification of APhPs in Au/$\alpha$-MoO$_3$/Au structure. (**a**) Schematic of near-field excitations of the APhPs nanoresonator. A thin Au square (width $w = 20$ nm and thickness $d = 5$ nm) is over the $\alpha$-MoO$_3$/Au layers, and the air gap is 0.5 nm between each layer. A vertical electric dipole (1 nm above the Au square) mimicking the tip of s-SNOM is used to excite and probe the local resonances of the nanoresonator. (**b**) Normalized decay rates $\Gamma/\Gamma_0$ ($\Gamma$ and $\Gamma_0$ are the decay rate of the dipole source for exciting APhPs and the decay rate of the dipole source in vacuum, respectively). M1 to M5 mark the resonance peak found in the spectra, corresponding to different orders of Fabry–Pérot resonances in the APhPs nanoresonator. (**c-e**) Normalized electric field distribution of the modes M1 to M3, taken from the plane at the height of 0.5 nm above the 2D $\alpha$-MoO$_3$ sheet.

For an experimental demonstration, one could use scattering-type scanning near-field optical microscope (s-SNOM) to excite and directly image the APhPs in the $\alpha$-MoO$_3$/metal heterostructure. However, for the metal/$\alpha$-MoO$_3$/metal sandwiched structure, the s-SNOM tip cannot directly access and image the APhPs as the top Au layer shields the electromagnetic fields of the APhPs. As illustrated in Fig. 4a, we propose a scheme for the experimental verification. We consider a thin Au square (width $w = 20$ nm and thickness $d = 5$ nm) over the $\alpha$-MoO$_3$/Au layers (the air gap is 0.5 nm between each layer), forming a nanoresonator of APhPs. We perform numerical simulations using a vertical electric dipole (1 nm above the Au square, mimicking the tip of s-SNOM) to excite and probe the local resonances of the nanoresonator. We calculate the decay rate of the dipole source ($\Gamma = -2\pi c^{-1}\text{Re}[E_z(\mathbf{r}_0)]$) for characterizing the electromagnetic interaction between the dipole and the APhPs resonator[52], as shown in Fig. 5b. We find five resonance peaks in the calculated spectra, corresponding to different order Fabry–Pérot resonances of APhPs in the nanostructure. The electric field distributions of the first three order modes are shown in Fig. 4c to e. We can see that the APhPs form the resonant patterns along the $x$-direction due to the canalization propagation. Moreover, one could also perform far-field experiments to probe the resonance peaks for the verification of the APhPs in the sandwich structure, as recently shown in the far-field experiments for detecting acoustic graphene plasmons[52,53].

## Conclusions

In summary, we theoretically propose and systematically explore in-plane anisotropic APhPs based on metal/MoO$_3$/metal sandwiched structures. Compared to the APhPs reported in other previous works[18,44], the APhPs in the sandwiched structure show the strongest electromagnetic-field confinement, the largest near-field enhancement, and can propagate as a canalization (directional, diffraction-less) waveguide mode along the layers. Furthermore, for a double layer of $\alpha$-MoO$_3$ sandwiched by two Au layers, the coupling of the APhPs modes can be adjusted by rotating the twist angle $\varphi$ between the optical axes of the two layers of $\alpha$-MoO$_3$. Finally, we also propose the APhPs-based nanoresonators for an experimental verification of the APhPs in the sandwiched structure. Our results show that the in-plane anisotropic APhPs based on the metal-MoO$_3$-metal sandwiched structures could be a promising platform for light guiding and manipulation at a deeply subwavelength scale.

## Conflicts of interest

There are no conflicts to declare.

## Acknowledgements

We acknowledge the start-up funding from Huazhong University of Science and Technology; and National Natural Science Foundation of China (Grant No. 62075070).